\documentclass[aps,prl,reprint,groupedaddress]{revtex4-2}

\usepackage{graphicx}

\usepackage{multirow}
\usepackage{appendix}

\begin{document}


\title{Simultaneous probes of pseudogap and disorder by hard x-ray photoemission applied for a candidate thermoelectric Al-Pd-Ru quasicrystal}


\author{
    N. U. Sakamoto$^{1, 2}$, G. Nozue$^{1, 2}$, H. Fujiwara$^{1, 2, 3}$, Y. Torii$^1$, M. Sakaguchi$^{1, 2}$, T. D. Nakamura$^{1, 2}$,\\
    T. Kiss$^1$, H. Sugawara$^4$, S. Tanaka$^5$, Y. Iwasaki$^6$, Y. Niwa$^5$, A. Ishikawa$^5$,\\
    T. D. Yamamoto$^5$, R. Tamura$^5$, A. Yasui$^7$, and A. Sekiyama$^{1, 2, 3}$
}
\affiliation{
    $^1$Division of Materials Physics, Graduate School of Engineering Science, The University of Osaka, Toyonaka, Osaka 560-8531, Japan\\
    $^2$RIKEN SPring-8 Center, Sayo, Hyogo 679-5148, Japan\\
    $^3$Spintronics Research Network Division, Institute for Open and Transdisciplinary Research Initiatives, The University of Osaka, Suita, Osaka 565-0871, Japan\\
    $^4$Depertment of Physics, Kobe University, Kobe, Hyogo 657-8501, Japan\\
    $^5$Department of Materials Science and Technology, Tokyo University of Science, Katsushika, Tokyo 125-8585, Japan\\
    $^6$National Institute for Materials Science, Tsukuba, Ibaraki 305-0047, Japan\\
    $^7$Japan Synchrotron Radiation Research Institute, Sayo, Hyogo 679-5198, Japan
    }


\date{\today}

\begin{abstract}
The bulk electronic structure of Al-Pd-Ru quasicrystal (QC) have been investigated by hard X-ray photoemission spectroscopy (HAXPES).
We have found an intrinsic pseudogap structure in which the spectral weight in the vicinity of the Fermi level ($E_F$) is remarkably suppressed at any photon energy.
The valence-band HAXPES spectra and the asymmetry of the core-level peaks indicate a contribution of the Al sites to the electronic structure in the vicinity of $E_F$ with less contributions from the Pd and Ru sites.
The disorder effects are found in the core-level lineshapes of the Al-Pd-Ru QC, of which the widths are much broader than those of the other reference crystalline solids with less disorder.
\end{abstract}


\maketitle

\section{I. Introduction}

Al-based quasicrystals (QCs), a group that has long been studied since the discovery of QCs \cite{Shechtman}, exhibit non-metallic behavior, such as high electrical resistivity \cite{akiyama1993toward,Fisher01062002}.
This property is attributed to the deep pseudogap structure formed in the vicinity of the Fermi level ($E_F$) by the Hume-Rothery mechanism \cite{HumuRothery,tsai2004test} and covalent bonding \cite{kirihara2000covalency,kirihara2003covalent}.
The Al-based QCs can exhibit high Seebeck coefficient $S$ since the suppression of the density of states (DOS) and steep changes in DOS due to the deep pseudogap structures can improve $S$.
In addition, the Al-based QCs exhibit low thermal conductivity comparable to that of glass due to its complex structure \cite{chernikov1995low}.
Attempts have been made to improve the thermoelectric performance of the Al-based QCs through compositional changes and/or elemental substitutions \cite{kirihara2002composition,nagata2003effect,takagiwa2008thermoelectric}.

So far, QCs and several approximant crystals (ACs) phases have been discovered in the Al-Pd-Ru system \cite{pavlyuchkov2008rich,pavlyuchkov2009investigation}, where it has been found that the electrical resistivity and its temperature dependence exhibit non-metallic behavior in QC and 2/1 AC phases \cite{tamura1999ordered}.
Recently, the largest power factor among quasicrystals has been obtained in Al-Pd-Ru QCs with a power factor $S^2\sigma$ of $780\,\mathrm{\mu W/(m\cdot K^2)}$
\footnote{Y. Iwasaki, unpublished.},
where $\sigma$ stands for the electrical conductivity.

Investigation of the electronic structure of the Al-based QCs is important to improve their thermoelectric property.
Since QCs do not have translational symmetry, band-structure calculations are very difficult to be carried out at present.
Therefore, photoemission spectroscopy (PES) is powerful for clarifying the electronic states near $E_F$ for QCs and ACs.
Stadnik \textit{et al.}~have performed high-resolution PES for various systems and have observed decreases in spectral weight at $E_F$ \cite{PhysRevB.55.10938}.
In some cases under surface-sensitive conditions, clear Fermi edges were observed for some Al-based QCs, pointing to a possible difference between the bulk and surface electronic structures of QCs \cite{PhysRevB.58.734}.
Bulk-sensitive measurement are more suitable to investigate the relationship between thermoelectric properties and electronic structure.
Hard X-ray photoemission spectroscopy (HAXPES) can probe the bulk electronic structure of solids \cite{tanuma2011calculations, sekiyama2016high}, by which the observations of pseudogap structures have been reported in Al-Pd-TM (TM: transition metal) systems\cite{PhysRevLett.109.216403,PhysRevResearch.3.013151,PhysRevB.103.L241106}.

To investigate the bulk electronic structure of the Al-Pd-Ru QC, we have performed valence-band and core-level HAXPES.
On the other hand, HAXPES spectra of materials including light elements such as Al can be affected by the recoil effets \cite{recoilYT,suga2010soft}.
We have confirmed the influence of the recoil effects by the HAXPES with changing excitation photon energy, and observed an intrinsic pseudogap structure of the Al-Pd-Ru QC.
We have also performed polarization-dependent valence-band and core-level HAXPES to reveal the electronic structure in the vicinity of $E_F$ orbital- and element-selectively.
These results have revealed the contribution of the Al sites to the DOS in the vicinity of $E_F$.
We have also observed a broadening of the core-level lineshapes of the Al-Pd-Ru QC compared with those of conventional crystalline solids, which reflects the chemical disorders on the sites in the Al-Pd-Ru QC \cite{hatakeyama2017atomic,fujita2013cluster}.
It should be noted that the disorder is indispensable for obtaining the freedom of composition ratio which leads to the various interesting phenomena for QCs and ACs.
The orbital- and element-dependent electronic states have been revealed by our HAXPES with changing the photon energy and polarization, which has unfortunately been overlooked in the previous HAXPES studies of QCs and ACs with the deep pseudogaps.

\section{II. EXPERIMENTAL}

Polycrystalline sample of $\mathrm{Al_{71.5}Pd_{19}Ru_{9.5}}$ QC was prepared as follows.
High-purity raw material powders Al (4N), Pd (4N), and Ru (3N) were weighed and mixed to achieve the stoichiometric composition.
The mixed high-purity powders were compacted into pellets using a hydraulic press.
The compacted pellets were melted in an arc melting furnace under an Ar atmosphere.
The ingot produced by arc melting was wrapped in tantalum foil, sealed in a quartz tube under an Ar atmosphere, and annealed in a muffle electric furnace at 1273 K for 65 hours.
After the heat treatment, the sample was cooled by water-quench.
The annealed ingot was crushed into powder using a mortar, and the resulting powder was sintered by spark plasma sintering (LABOX-110MC; SinterLand, Inc) under an Ar atmosphere at a uniaxial pressure of 90 MPa, at approximately 1173 K for $5-10$ minutes.

The HAXPES experiments for the Al-Pd-Ru QC were performed at BL09XU in SPring-8 \cite{bl09xu} with SCIENTA OMICRON R4000 photoelectron spectrometer.
The photon energy was set to 5.0 keV, 7.2 keV, and 9.5 keV for the valence-band HAXPES\@.
A Si(111) double-crystal monochromator selected linearly polarized within the horizontal plane, which was further monochromatized by the double channel-cut crystals using a Si(220) reflection for $h\nu=$ 5.0 keV and Si(311) reflections for $h\nu=$ 7.2 and 9.5 keV\@. 
At $h\nu=$ 5.0 keV and 9.5 keV, the HAXPES were performed with the horizontal linearly polarized X-ray, while at $h\nu=$ 7.2 keV we switched between horizontaly and verticaly linear polarizations with a diamond phase retarder.
The degrees of linear polarization ($P_L$) for the horizontaly and verticaly polarized X-rays were $\sim\!\!0.94$ and $-0.92$ at $h\nu=$ 7.2 keV.
The core-level HAXPES was performed at $h\nu=$ 7.2 keV\@.
The overall energy resolution was set to 180 (200) meV at $h\nu=$ 7.2 keV (5.0 keV and 9.5 keV)\@.
Note that such an energy resolution, which is better in the practical HAXPES, is required for the core-level line-shape analyses discussed below in addition to the valence-band measurements.
The binding energy was calibrated by measuring the Fermi edge of the Au metal electrically connected to the Al-Pd-Ru QC before and after the valence-band HAXPES measurements.
The clean surface was obtained by fracturing \textit{in situ} at the measuring temperature of 40 K, at which we can assume that additional temperature-dependent broadening of the core-level HAXPES spectra by phonons (except for the recoil) is negligible relative to the lifetime broadening and the experimental energy resolutions.
We also performed the HAXPES at BL19LXU in SPring-8 \cite{Fujiwara:ve5047} for single-crystalline $\mathrm{NdTi_2Al_{20}}$ and $\mathrm{CeRu_2Ge_2}$ to compare the asymmetry and widths of the core-level main peaks as well as the spectral weights in the vicinity of $E_F$.
The photon energy was set to $\sim\!\!8$ keV\@.
The energy resolution was set to 410 (150) meV for Al $1s$ core-level and valence-band (Al $2p$ core-level) HAXPES for $\mathrm{NdTi_2Al_{20}}$.
For the HAXPES of $\mathrm{CeRu_2Ge_2}$, the experimental coonditions are discribed elsewhere \cite{PhysRevB.77.035118}.

\section{III. RESULTS AND DISCUSSIONS}

\subsection{A. Photon-energy-dependent valence-band HAXPES}

\begin{figure*}
    \includegraphics{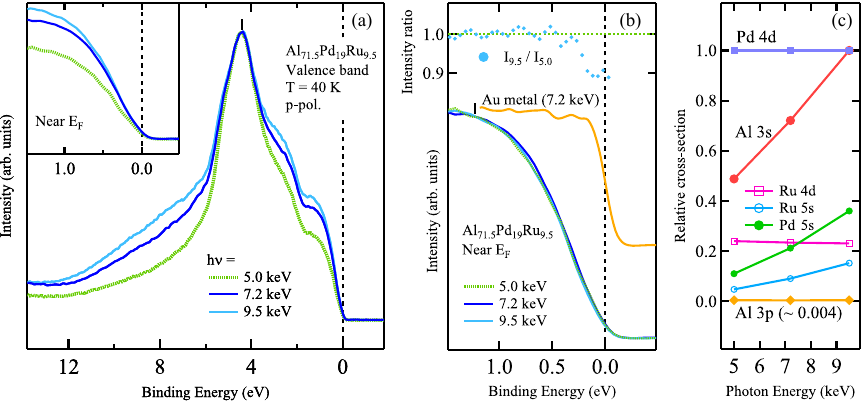}
    \caption{\label{VB_hv}
    (a) Photon-energy dependence of the valence-band HAXPES spectra of the Al-Pd-Ru QC\@.
    Those in the near-$E_F$ region is shown in the inset.
    The spectrum measured at $h\nu=$ 7.2 keV has been broadened by a Gaussian to have the same energy resolution as that of the spectra at $h\nu=$ 5.0 and 9.5 keV.
    (b)(bottom) Comparison of the valence-band HAXPES spectra near $E_F$ between the Al-Pd-Ru QC and Au metal.
    The spectra were normalized by the spectral weight around (a) 4.5 eV and (b) 1.2 eV, respectively, as indicated by black lines.
    (top) Intensity ratio obtained by dividing the spectral weight at $h\nu=$ 9.5 keV by the spectral weight at $h\nu=$ 5.0 keV\@.
    (c) Relative photoionization cross-sections for the p-pol.~configration \cite{trzhaskovskaya2018dirac} as a function of photon energy considering the composition of the Al-Pd-Ru QC.
    The relative cross-sections for Pd $5s$ orbital have been obtained by using the parameter of the Ag $5s$ orbital because of no calculation for the Pd $5s$ orbital.
    The number of electrons occupying each orbital was assumed to be the number shown in Table \ref{Ele_num}.
    }
\end{figure*}

Figure \ref{VB_hv}(a) shows the valence-band HAXPES spectra of the Al-Pd-Ru QC measured with 5.0 keV, 7.2 keV, and 9.5 keV photon energies at the p-polarization (p-pol.)~configuration.
There are peaks around 4.5 eV and shoulder structures around 2.5 and 1.2 eV in all spectra.
These features are similar to those seen in the HAXPES spectra of other Al-Pd-TM systems \cite{PhysRevLett.109.216403,PhysRevResearch.3.013151,PhysRevB.103.L241106}.
As the photon energy increases, the structures around $0-3.5$ eV and $6-12$ eV are enhanced relative to the peaks around 4.5 eV\@.
The spectral weights near $E_F$ show steep changes as a function of binding energy and are strongly suppressed at any photon energy. 

To investigate the electronic structure in the vicinity of $E_F$ as well as the influence of the recoil effects\cite{recoilYT,suga2010soft}, the spectra have been normalized by the spectral weight around 1.2 eV as shown in Fig.~\ref{VB_hv}(b).
The spectral weights of the Al-Pd-Ru QC are suppressed in the vicinity of $E_F$ compared to that of conventional metal Au.
The Fermi cut-off is not observed in the spectra of the Al-Pd-Ru QC at any photon energy while a finite spectral weight seems to be seen at $E_F$.
Decrease of the intensity ratio $I_{9.5}/I_{5.0}$ for $h\nu=$ 9.5 keV ($I_{9.5}$) and 5.0 keV ($I_{5.0}$) in the vicinity of $E_F$ indicates that the spectral weight at $h\nu=$ 9.5 keV is more suppressed than that at $h\nu=$ 5.0 keV from $E_F$ to the binding energy of $\sim\!\!200$ meV, where the energy scale is comparable to the energy loss due to the recoil effects at the Al sites in the single-atom model as 190 meV at $h\nu=$ 9.5 keV.
This suppression would be due to the recoil effects, suggesting that the electrons in the Al sites contribute to the electronic structure in the vicinity of $E_F$.
Although the recoil effects slightly affect the valence-band HAXPES spectra of the Al-Pd-Ru QC, the suppression of the spectral weight in the vicinity of $E_F$ is essentially due to the intristic pseudogap structure.

When we take the composition of the Al-Pd-Ru QC and the preliminarily assumed valence electron number on each site shown in Table \ref{Ele_num} into account, it is noticed that the Pd $4d$ orbitals have the largest photoionization cross-section at $h\nu=$ 7.2 keV
\cite{trzhaskovskaya2018dirac}.
The calculated relative photoionization cross-sections to the Pd $4d$ orbitals at each photon energy are shown in Fig.~\ref{VB_hv}(c).
We can notice from the figure that the Pd $4d$ and Al $3s$ contributions are dominant while the contribution of the Al $3p$ orbitals is negligible at any photon energy.
The ratio of the contribution of the Ru $4d$ orbitals to that of the Pd $4d$ orbitals is nearly independent of photon energy whereas the contributions of $s$ orbitals increase relative to that of the Pd $4d$ orbitals with increasing photon energy.
The peak around 4.5 eV is ascribed to the Pd $4d$ contributions, which has been verified by the polarization dependence discussed below.
The photon-energy dependence of the relative photoionization cross-sections suggests that the evolutions of the spectral weights in $E_F$ to 3.5 eV and in $6-12$ eV in Fig.~\ref{VB_hv}(a) are due to the enhancement of the $s$-orbitals contribution at the higher excitation photon energy. 

\begin{table}
    \caption{\label{Ele_num}
    Number of electrons occupied in each orbital assumed for calculation of the relative photoionization cross-sections in Fig.~\ref{VB_hv}(c).
    }
    \begin{ruledtabular}
    \begin{tabular}{cccccc}
        \multicolumn{6}{c}{Number of electrons}\\
        \hline
        Al $3s$ & Al $3p$ & Pd $4d$ & Pd $5s$ & Ru $4d$ & Ru $5s$\\
        \hline
        2 & 1 & 10 & 1 & 7 & 1\\
    \end{tabular}
    \end{ruledtabular}
\end{table}

\subsection{B. Polarization-dependent valence-band HAXPES}

\begin{figure}
    \includegraphics{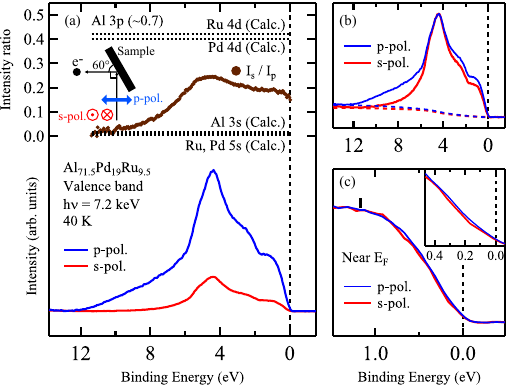}
    \caption{\label{VB_sp}
    (a) (bottom) Linear-polarization dependence of the valence-band HAXPES spectra of the Al-Pd-Ru QC at $h\nu=$ 7.2 keV\@.
    The Shirley-type backgrounds were subtracted from the raw spectra [see (b)].
    The spectral weights have been normalized by the photon flux.
    (top) Intensity ratio $I_s/I_p$ obtained by dividing the spectral weight at the s-pol.~configuration by that at the p-pol.~configuration.
    The dashed lines indicate the calculated $I_s/I_p$ for the Al $3s$, Pd $4d$, Ru $4d$, Ru $5s$, and Pd $5s$ states \cite{trzhaskovskaya2018dirac}.
    The $I_s/I_p$ for Pd $5s$ state has been obtained by using the parameter of the Ag $5s$ state because of no calculation for the Pd $5s$ state.
    The insert shows the experimental geometry.
    (b) Raw HAXPES spectra at $h\nu=$ 7.2 keV\@ before the background subtractions.
    The dashed lines indicate the Shirley-type backgrounds.
    The spectra were normalized by the spectral weight around 4.5 eV\@.
    (c) Comparison of the spectra at the different polarization configurations at $h\nu=$ 7.2 keV, which have been normalized at the spectral weight at 1.2 eV indicated by the black solid bar.
    The inset shows the same but in an expanded energy scale.
    }
\end{figure}

\begin{table}
    \caption{\label{isip}
    Calculated spectral intensity ratio $I_s/I_p$ for Al $3s$, Al $3p$, Pd $4d$, Ru $4d$, Ru $5s$, and Pd $5s$ orbitals at $h\nu=$ 7.2 keV\cite{trzhaskovskaya2018dirac}.
    }
    \begin{ruledtabular}
    \begin{tabular}{cccccc}
        \multicolumn{6}{c}{$I_s/I_p$}\\
        \hline
        Al $3s$ & Al $3p$ & Pd $4d$ & Pd $5s$ & Ru $4d$ & Ru $5s$\\
        \hline
        0.02 & 0.73 & 0.40 & 0.01 & 0.42 & 0.01\\
    \end{tabular}
    \end{ruledtabular}
\end{table}

Polarization dependence of the valence-band HAXPES spectra provides an orbital-selective electronic structure of the valence bands
\cite{sekiyama2010prominent}.
Figure \ref{VB_sp}(a) shows the valence-band HAXPES spectra of the Al-Pd-Ru QC measured at the p-pol.~and s-polarization (s-pol.)~configurations at $h\nu=$ 7.2 keV after subtracting the backgrounds shown in Fig.~\ref{VB_sp}(b).
The experimental spectral intensity ratio $I_s/I_p$ for the s-pol.~($I_s$) and the p-pol.~($I_p$) compared with the calculated $I_s/I_p$ for the atomic orbitals forming the valence bands \cite{trzhaskovskaya2018dirac} is also shown in the figure and Teble \ref{isip}.
Although the calculated $I_s/I_p$ of the Al $3p$ orbitals is the largest among those of the other orbitals in the Al-Pd-Ru QC, the Al $3p$ contribution orbitals is negligible even at the s-pol.~configuration due to the relatively weakest photoionization cross-section shown in Fig.~\ref{VB_hv}(c).
Since the calculated $I_s/I_p$ of the $s$ orbitals is very small ($<0.02$), the Pd and Ru $4d$ contributions are dominant in the spectrum at the s-pol.~configuration.
From the fact that Pd accounts for twice the composition of Ru, the peaks around 4.5 eV are ascribed to the Pd $4d$ orbitals.
The shoulder structures at lower binding energy side than the Pd $4d$ peaks are reduced in the valence-band HAXPES spectra of Al-Pd-3d TM systems \cite{PhysRevResearch.3.013151} and enhanced in those of Al-Pd-5d TM systems \cite{PhysRevB.103.L241106}.
This difference would be due to the difference in the photoionization cross-sections of the TMs \cite{trzhaskovskaya2018dirac}.
Therefore, the shoulder structures around 2.5 and 1.2 eV in the valence-band HAXPES spectra of the Al-Pd-Ru QC is ascribed to Ru $4d$ orbitals at the s-pol.~configuration.
As discussed above based on the $h\nu$ dependence of the HAXPES spectra at the p-pol.~configuration, the $s$ orbitals contribute also to these shoulder structures. 

The gradually decreasing experimental $I_s/I_p$ from 4.5 eV towards $E_F$ is rather abruptly reduced in the vicinity of $E_F$, which is consistent with the fact that the spectral weight at the p-pol.~configuration slightly exceeds that at the s-pol.~configuration as shown in Fig.~\ref{VB_sp}(c).
This indicates that the $s$-orbitals contribution is relatively large in the vicinity of $E_F$.
These $s$ contributions are mainly originating from the Al $3s$ orbitals based on the possible influence of the recoil effects [see Fig.~\ref{VB_hv}(b)] discussed above, and the following discussion based on the core-level HAXPES spectra. It is found that the pseudogap structure in the vicinity of $E_F$ is robust against the polarization and photon energy in the spectra of the Al-Pd-Ru QC.

\subsection{C. Asymmetry of core-level peaks}

\begin{figure}
    \includegraphics{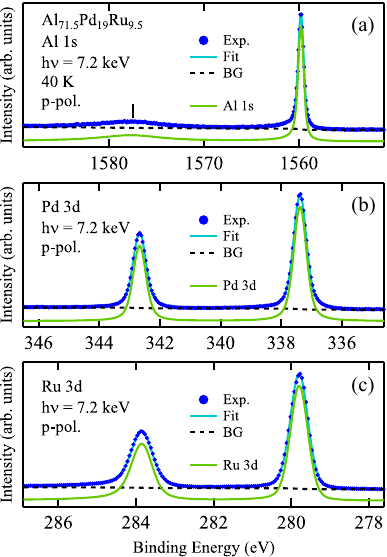}
    \caption{\label{Core_exp_fit}
    (a) Al $1s$, (b) Pd $3d$, and (c) Ru $3d$ core-level HAXPES spectra of the Al-Pd-Ru QC\@.
    The spectra were fitted by the combination of the Doniach-\v{S}unji\'{c} \cite{doniach1970many} line shapes broadened by a Gaussian shown by the green lines and the Shirley-type backgrounds.
    The dashed lines indicate the background of each spectrum.
    The black solid bar in (a) indicates the plasmon peak.
    }
\end{figure}

Figure \ref{Core_exp_fit} shows the core-level HAXPES spectra of the Al-Pd-Ru QC\@.
According to our line-shape analyses \cite{doniach1970many}, the sharp Al $1s$ main peak is located at $\sim\!\!1559.8$ eV and the broad plasmon satellite is seen at the higher binding energy by $\sim\!\!17.7$ eV from the main peak.
This peak binding energy is slightly higher by $\sim\!\!0.3$ eV than that for Al metal ($\sim\!\!1559.5$ eV) \cite{suga2010soft}.
The Pd $3d_{5/2}$ and $3d_{3/2}$ peaks of the Al-Pd-Ru QC are located at $\sim\!\!337.4$ eV and $\sim\!\!342.7$ eV, respectively.
These peaks are shifted to the higher binding energy side, about 2.4 eV higher than those of the Pd metal \cite{4dTM}.
On the other hand, the Ru $3d_{5/2}$ and $3d_{3/2}$ peaks are located at $\sim\!\!279.8$ eV and $\sim\!\!283.8$ eV, respectively, shifting toward the lower binding energy side by about 0.3 eV compared to the Ru metal \cite{4dTM}.
Although the straightforward discussion would be rather premature, the different tendency of the peak shifts relative to the spectra of the simple metals implies that the valence/conduction electrons are transferred from the Ru to Al, Pd sites if their interaction with the created core holes could be negligible and hence the core levels would be determined by the rigid-band-like shifts due to the charge transfer.

The electronic structure in the vicinity of $E_F$ is also reflected in the line shapes of core-level peaks.
Core-level peaks of metals tail asymmetrically towards the high binding energy side due to the energy loss from electron-hole pair excitations in the vicinity of the $E_F$ \cite{doniach1970many}.
The large partial density of states (PDOS) at $E_F$ is expected to increase the asymmetry of the core-level peaks on the sites since the probability of the electron-hole pair formations is larger for the larger PDOS.
For the Al-Pd-Mn QC, Neuhold \textit{et al.}~\cite{PhysRevB.58.734} and Horn \textit{et al.}~\cite{horn2005core} have shown that the Al-Pd-Mn QCs have the symmetric core-level peaks compared to those of single metals due to the pseudogap structure near $E_F$.

\begin{figure}
    \includegraphics{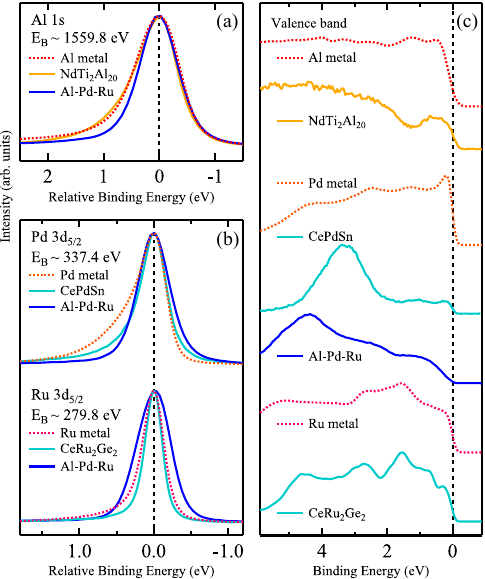}
    \caption{\label{Core_asy_VB}
    (a) Comparison of the Al $1s$ core-level PES spectrum of the Al-Pd-Ru QC with those of the reference metallic materials Al metal \cite{suga2010soft} and $\mathrm{NdTi_2Al_{20}}$.
    (b) Pd $3d_{5/2}$ and Ru $3d_{5/2}$ core-level PES spectra of the Al-Pd-Ru QC compared with those of the metallic Pd, Ru metals \cite{4dTM}, CePdSn \cite{sekiyama1999high,sekiyama2005high}, and $\mathrm{CeRu_2Ge_2}$ \cite{PhysRevB.77.035118}.
    The abscissas are the binding energy relative to the peak-top binding energy.
    The Shirley-type backgrounds have been subtracted from the raw spectra.
    To have the mutually similar energy resolutions, the spectra with the better energy resolution have been broadened by a Gaussian.
    (c) Comparison of the valence-band PES spectrum of the Al-Pd-Ru QC ($h\nu=$ 7.2 keV at the p-pol.~configuration) with those of the reference metallic materials.
    Note that the Fermi edge of the Al metal is shifted due to the recoil effects.
    }
\end{figure}

We have compared the core-level line shapes of the Al-Pd-Ru QC with those of the reference metallic materials, as shown in Fig.~\ref{Core_asy_VB}.
It is clear that there is a correlation between the PDOS at $E_F$ and the asymmetry of core-level peaks.
For instance, the Al $1s$ core-level HAXPES spectra clearly show the asymmetric tails for the Al metal and $\mathrm{NdTi_2Al_{20}}$ whereas the Fermi edges are seen in the their valence-band spectra as shown in Figs.~\ref{Core_asy_VB}(a,c).
On the other hand, the Al $1s$ peak seems to be rather symmetric for the Al-Pd-Ru QC.
The core-level peak of Pd metal is prominently asymmetric reflecting the large DOS at $E_F$.
CePdSn, of which the Pd $4d$-dominant soft x-ray valence-band spectrum shows the metallic Fermi edge but the PDOS is relatively reduced compared with that  for the Pd metal, shows the less asymmetric Pd $3d$ core-level peaks than those of the Pd metal as shown in Figs.~\ref{Core_asy_VB}(b,c).
The spectral weight in the vicinity of $E_F$ for the Al-Pd-Ru QC is strongly suppressed and its core-level peaks is symmetric compared to those of the Pd metal and CePdSn.
For the Ru sites, the asymmetric tail is seen in the relative binding energy region of $0.5-1.5$ eV for the Ru metal of which the Fermi edge is seen in the valence-band HAXPES spectrum as shown in Figs.~\ref{Core_asy_VB}(b,c).
The core-level peaks of the Al-Pd-Ru QC are symmetric compared to those of the reference metallic materials showing the asymmetric tails; the contribution to the PDOS in the vicinity of $E_F$ is relatively weak at all sites of the Al-Pd-Ru QC as shown in Fig.~\ref{Core_asy_VB}(c).

\begin{figure}
    \includegraphics{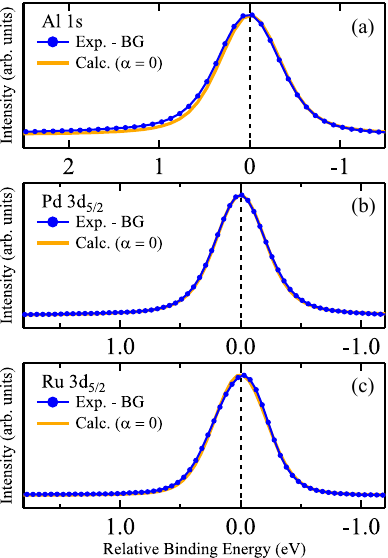}
    \caption{\label{Core_asy_a0}
    (a) Al $1s$, (b) Pd $3d_{5/2}$, and (c) Ru $3d_{5/2}$ core-level peaks of the Al-Pd-Ru QC compared with the simulated symmetric peaks with the same peak widths of those of the Al-Pd-Ru QC obtained by the fitting.
    }
\end{figure}

To quantify the asymmetry of the core-level peaks, we have compared singularity index $\alpha$ obtained by the fitting using the Doniach-\v{S}unji\'{c} function \cite{doniach1970many} as shown in Fig.~\ref{Core_exp_fit} and Fig.~S1 in the Supplemental Material \cite{supplemental}.
Note that the asymmetry becomes larger for larger $\alpha$
\footnote{The Doniach-\v{S}unji\'{c} function has an advantage in representing the asymmetry by using the unique parameter $\alpha$ compare with other proposed functions (for example, the formula given by Mahan \cite{mahan1975collective}).
Since we evaluate only asymmetry by the fitting, the difficulty in the area divergence of the Doniach-\v{S}unji\'{c} funtion is not serious for our analyses.}.
Table \ref{Core_alpha_fwhm} shows the obtained $\alpha$ values of the core-level peaks of the Al-Pd-Ru QC and reference materials.
It should be noted that the Pd $3d_{5/2}$ peak of Pd metal could not be fitted well because the Doniach-\v{S}unji\'{c} function implicitly assumes that the PDOS changes hardly near $E_F$ \cite{GKWertheim_1976}.
The Pd metal has the PDOS that is inconsistent with this assumption as shown in Fig.~\ref{Core_asy_VB}(c).
The singularity index $\alpha$ has been uniquely evaluated for all core-level peaks except for those of the Pd metal, since the line shapes of the other materials are well reproduced.
The $\alpha$ values for the Al-Pd-Ru QC are remarkably smaller than those for the reference metallic materials at all sites.
For the Al-Pd-Ru QC, the Al $1s$ peak has the larger $\alpha$ than those for the Pd and Ru $3d_{5/2}$ peaks.
Indeed, the comparison of the experimental main peaks with the simulated symmetric ones shown in Fig.~\ref{Core_asy_a0} tells us that only the Al $1s$ core-level main peak shows a slightly asymmetric tail while the Pd and Ru $3d_{5/2}$ main peaks are nearly symmetric.
These results imply that the Al sites seem to contribute to the DOS in the vicinity of $E_F$.
The $\alpha$ value of the Pd $3d_{5/2}$ peak is an order of magnitude smaller than those of the Al $1s$ and Ru $3d_{5/2}$ peaks, from which we conclude that the contribution of the Pd sites to the DOS near $E_F$ is relatively small.
The Ru $3d_{5/2}$ peak is also nearly symmetric, however, the $\alpha$ value is larger than that of the Pd $3d_{5/2}$ peak, which could suggest a slight contribution of the Ru sites to the conduction electrons.
These results are consistent with those of the valence-band HAXPES discussed above and support the pseudogap structure of the Al-Pd-Ru QC\@.

\begin{table}
    \caption{\label{Core_alpha_fwhm}
    Singularity index $\alpha$ giving the asymmetric line shapes of the core-level peaks for the Al-Pd-Ru QC and reference materials, and full width at half maximum (FWHM) of a Gaussian used for broadening the spectra to reproduce the spectra of the lower binding energy side of the main peak for the Al-Pd-Ru QC as shown in Fig.~\ref{Core_wid}.
    The $\alpha$ values are shown with estimated error.
    The term "poly" ("single") stands for a polycrystalline (single-crystalline) sample.
    }
    \begin{ruledtabular}
    \begin{tabular}{cccc}
        \multirow{2}{*}{\textrm{Al}}&
        Al-Pd-Ru QC & Al metal & $\mathrm{NdTi_2Al_{20}}$\\
        & Poly & Poly & Single\\
        \cline{2-4}
        $\alpha$ (from $1s$) & $0.040\pm0.001$ & $0.124\pm0.002$ & $0.124\pm0.002$ \\
        \begin{tabular}{c}
        FWHM [eV]\\(from $2p$) 
        \end{tabular}& -- & 0.33 & 0.28 \\
        \hline
        \multirow{2}{*}{Pd $3d_{5/2}$}&
        Al-Pd-Ru QC & Pd metal & CePdSn\\
        & Poly & Poly & Poly\\
        \cline{2-4}
        $\alpha$ & $0.001\pm0.001$ & $0.185\pm0.005$ & $0.112\pm0.002$\\
        FWHM [eV] & -- & 0.31 & 0.29 \\
        \hline
        \multirow{2}{*}{Ru $3d_{5/2}$}&
        Al-Pd-Ru QC & Ru metal & $\mathrm{CeRu_2Ge_2}$\\
        & Poly & Poly & Single\\
        \cline{2-4}
        $\alpha$ & $0.018\pm0.002$ & $0.138\pm0.001$ & $0.072\pm0.002$ \\
        FWHM [eV] & -- & 0.36 & 0.41 \\
    \end{tabular}
    \end{ruledtabular}
\end{table}

\subsection{D. Disorder effects in the core-level photoemission spectra}

\begin{figure}
    \includegraphics{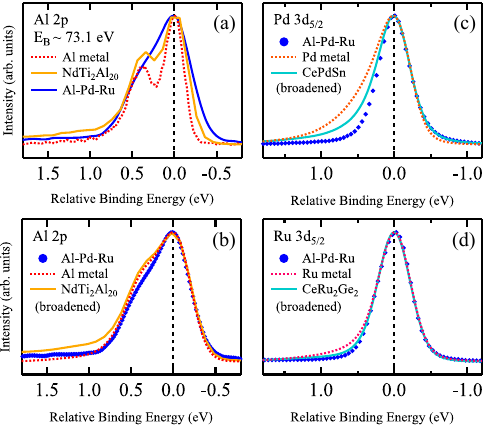}
    \caption{\label{Core_wid}
    (a) Al $2p$ core-level HAXPES spectrum of the Al-Pd-Ru QC measured with the s-pol.~configuration compared with the spectra of reference materials (Al metal \cite{suga2010soft} and $\mathrm{NdTi_2Al_{20}}$) after subtracting the Shirley-type backgrounds.
    (b,c,d) Broadened core-level photoemission spectra of the reference materials in Fig.~\ref{Core_asy_VB}(b) and (a).
    Each spectrum has been broadened by a Gaussian with the full width at half maximum of the value in Table \ref{Core_alpha_fwhm} so that the peak width on the low binding energy side well fits those of the Al-Pd-Ru QC\@.
    }
\end{figure}

The Pd and Ru $3d_{5/2}$ peaks of the Al-Pd-Ru QC are much broader than those of the reference materials as shown in Fig.~\ref{Core_asy_VB}(b).
The broadening effect for the Al sites has beeen compared using the Al $2p$ peaks as shown in Fig.~\ref{Core_wid}(a) because the Al $1s$ lifetime broadening is large as $\sim\!\!0.4$ eV while the natural width of the Al $2p$ orbitals is negligible \cite{krausewidth}.
Note that the Al $2p$ core-level HAXPES spectrum of the Al-Pd-Ru QC has been measured at the s-pol.~configuration to effectively reduce the influence of the broad Ru $4s$ peak of which the binding energy is only $\sim\!\!2$ eV higher than that of the Al $2p$ peaks (see Fig.~S2 in the Supplemental Material \cite{supplemental}).
The spin-orbit splitting of $\sim\!\!0.4$ eV is observed in the Al $2p$ core-level HAXPES spectra of the reference materials, which is not observed in that of the Al-Pd-Ru QC due to the remarkable broadening.

The spectral broadening has been evaluated by a Gaussian broadening.
Since the Pd and Ru $3d_{3/2}$ peaks would be relatively broadened due to $M_4M_5V$ ($3d_{3/2}$-$3d_{5/2}$-valence) Coster-Kronig transitions \cite{Coster,4dTM,PhysRevB.77.035118}, the comparison has been made with only the Pd and Ru $3d_{5/2}$ main peaks.
The high binding energy side of the core-level peaks is affected by the difference in the asymmetry; thus, the narrower peaks of the reference materials have been broadened to reproduce the spectral weight at the low binding energy side of the peaks of the Al-Pd-Ru QC as shown in Figs.~\ref{Core_wid}(b,c,d).
The optimized full width at half maximum (FWHM) values for the narrower peaks of the core levels for the reference materials are listed in Table \ref{Core_alpha_fwhm}.
The Ru $3d_{5/2}$ peak of $\mathrm{CeRu_2Ge_2}$ is sharp and therefore its FWHM value is larger than that of the Ru metal.
Although $\mathrm{NdTi_2Al_{20}}$ employed for the HAXPES measurements is single-crystalline, its Al $2p$ lineshape is more broadened than that of the polycrystalline Al metal.
This difference can be explained by the multi-site effect.
Since the photoemission spectra are formed by the superpositions of contributions from all sites, different binding energies at each site broaden the peak width.
$\mathrm{NdTi_2Al_{20}}$ has three crystallographically different Al sites \cite{sugawara} while the Al metal has one Al site.
The difference of the local environment surrounding the atoms at each Al site is reflected in the core-level peak width.
For the Pd $3d_{5/2}$ core-level peak, the FWHM values of $\sim\!\!0.3$ eV has been optimized for both polycrystalline Pd metal and CePdSn.
This value is rather comparable to that for the Al $2p$ peaks as well as for the Ru $3d_{5/2}$ peak while the broadening effect seems to be slightly larger for the Ru sites.

For the Al-Pd-Ru QC, in addition to the multiple sites, the disorders would cause the broadening of the core-level lineshapes.
The electrostatic (Madelung) potentials are slightly different between the mixed sites caused by the chemical disorders and the original sites, which leads to the slight site-dependent differences in the binding energy.
Then the core-level PES spectra are broadened by the superposition from all sites.
It should be noted that the clear broadening of the spectrum at the lower binding energy side of the Al $2p$ main peaks of the Al-Pd-Ru QC relative to that of $\mathrm{NdTi_2Al_{20}}$ with three Al sites cannot be explained solely by the multi-site effect.
Al-Pd-Ru QC is composed of two types of clusters: pseudo-Mackay clusters and mini-Bergman clusters, and there are chemical disorders with atom/ion mixing in several sites \cite{hatakeyama2017atomic}.
In the previous study, we have found that the peak widths change between rare-earth-based ACs with and without chemical disorders \cite{nozue2024high}.
The site dependence of the Madelung potential due to the chemical disorder effects gives the remarkable core-level spectral broadenings for the Al-Pd-Ru QC. 

It is known that the extra-atomic relaxation yielded by the screening of the conduction electrons can also affect the core-level lineshapes broadening \cite{cole1995extra}.
This effect would rather compensate the differences in the Madelung potential caused by the disorders, which reduces the broadening with enough conduction electrons.
The broadening effect in the Al-Pd-Ru QC is found to be comparable among the Al, Pd, and Ru sites, FWHM of $\sim\!\!0.3$ eV\@.
This means that there are not enough conduction electrons at any site of the Al-Pd-Ru QC for effectively giving the extra-atomic relaxation.
If some sites show the extra-atomic relaxation and others do not, there would be a marked difference in the broadening effect \cite{nozue2024high}.
In the photoemission process, the extra-atomic relaxation due to the screening can be categorized as the inter-site effect while the electron-hole-pair excitations leading to the asymmetric peak discussed above is regarded as the single-site effect.
Since the single-site effects are more likely to occur than the inter-site effects in the core-level photoemission process in which the core hole is suddenly created on the sites, the observation of the asymmetric Al $1s$ peak is not contradictory to the absence of the extra-atomic relaxation.
Namely, there are finite conducting carriers in the Al sites as revealed by Fig.~\ref{Core_asy_a0}(a) but the carrier density is not enough for the relaxation in the Al-Pd-Ru QC.  
This result also supports the intristic pseudogap structure of the Al-Pd-Ru QC.

\section{IV. SUMMARY}

We have performed the HAXPES of the Al-Pd-Ru QC with various photon energies and polarizations.
From the valence-band HAXPES, we have found the intristic pseudogap electronic structure of the bulk Al-Pd-Ru QC.
In addition, the analyses of the high-resolution core-level HAXPES spectra of the Al-Pd-Ru QC in comparison with those of the reference metallic materials provide the element- and orbital-selective electronic structure in the vicinity of $E_F$ and the chemical disorder effects.
The spectroscopic technique using high-resolution core-level and valence-band HAXPES employed here is applicable also for materials to simultaneously investigate the electronic structure and the disorder effects.

\section*{ACKNOWLEDGEMENTS}
\begin{acknowledgments}
    We acknowledge S. Yonezawa, Y. Murakami, S. Kaneko, A. Higashiya, A. Yamasaki, Y. Kanai-Nakata, S. Imada, S. Hamamoto, and K. Tamasaku for supporting the experiments.
    We also thank D. T. Adroja, T. Takabatake, and Y. Onuki for providing the high-quality CePdSn and $\mathrm{CeRu_2Ge_2}$ samples.
    The experiments in SPring-8 were performed under the approval of JASRI (Proposal Nos. 2024A1645, 2024B1899, and 2024B1957) whereas those at BL19LXU in SPring-8 were performed under the approval of JASRI and RIKEN (Proposal Nos. 2024A1398, 2024B1511, and 20240066).
    This work was financially supported by a Grant-in-Aid for Innovative Areas (JP20H05271 and JP22H04594), a Grant-in-Aid for Transformative Research (JP23H04867), a Grant-in-Aid for Scientific Research (JP22K03527 and JP24K03202) from JSPS and MEXT, Japan, and CREST (JPMJCR22O3) from JST, Japan. G. Nozue was supported by the Osaka University fellowship program of Super Hierarchical Materials Science Program and by the JSPS Research Fellowships for Young Scientists.
\end{acknowledgments}

\bibliography{AlPdRu_No_HAXPESforarXivv2}

\begin{widetext}

\clearpage

\appendix

\section*{Appendix: Supplementary Materials}

\begin{figure}
    \includegraphics{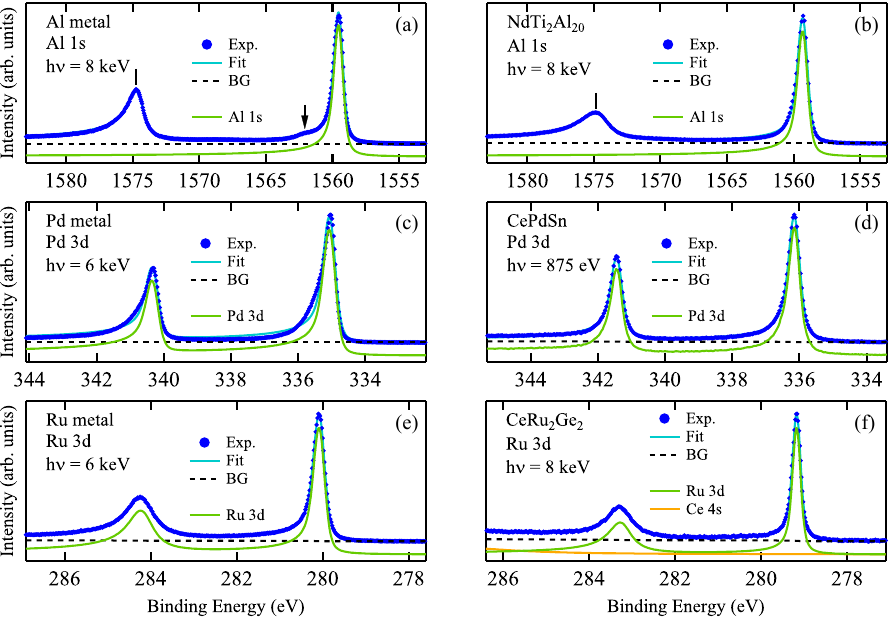}
    \caption{\label{ref_fit}
    (a,b) Al $1s$, (c,d) Pd $3d_{5/2}$, and (e,f) Ru $3d_{5/2}$ core-level photoemission spectra of the reference metallic materials (Al metal \cite{suga2010soft}, $\mathrm{NdTi_2Al_{20}}$, Pd and Ru metal \cite{4dTM}, CePdSn \cite{sekiyama1999high,sekiyama2005high}, and $\mathrm{CeRu_2Ge_2}$ \cite{PhysRevB.77.035118}).
    The spectra were fitted by the combination of the Doniach-\v{S}unji\'{c} \cite{doniach1970many} line shapes broadened by a Gaussian shown by the green lines and the Shirley-type backgrounds.
    The dashed line indicates the background of each spectrum.
    The black solid bar in (a,b) indicates the plasmon peak.
    The black arrow in (a) indicates the shoulder caused by the oxidation ($\sim\!\!1562.0$ eV)\@.
    }
\end{figure}

\begin{figure}
    \includegraphics{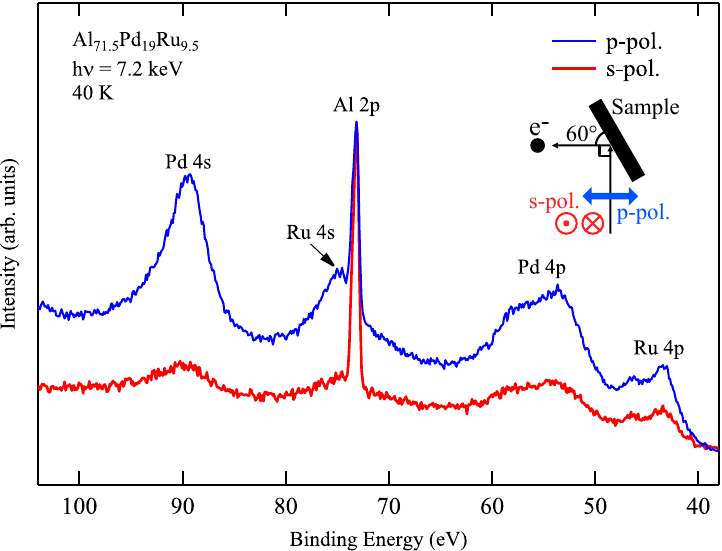}
    \caption{\label{Al2p_sp}
    Linear-polarization dependence of the Al $2p$, Pd $4s$, Pd $4p$, Ru $4s$, and Ru $4p$ core-level HAXPES spectra of the Al-Pd-Ru QC at $h\nu=$ 7.2 keV\@.
    The insert shows the experimental geometry.
    }
\end{figure}

\begin{table}
    \caption{\label{Al2p_isip}
    The calculated $I_s/I_p$ for the Al $2p$, Pd $4s$, Pd $4p$, Ru $4s$, and Ru $4p$ states \cite{trzhaskovskaya2018dirac}.
    }
    \begin{ruledtabular}
    \begin{tabular}{cccccc}
        orbital & Al $2p$ & Pd $4s$ & Pd $4p$ & Ru $4s$ & Ru $4p$\\
        \hline
        $I_s/I_p$ & 0.71 & 0.01 & 0.14 & 0.01 & 0.15
    \end{tabular}
    \end{ruledtabular}
\end{table}

\end{widetext}

\end{document}